    \documentclass[conference]{IEEEtran}
    
    \usepackage{cite}
    \usepackage{amsmath,amssymb,amsfonts}
    \usepackage{algorithmic}
    \usepackage{graphicx}
    \usepackage{textcomp}
    \usepackage{xcolor}
    \usepackage{hyperref}
    \usepackage[utf8]{inputenc}
    \usepackage{pgfplots}
    \usepackage{subcaption}
    \DeclareUnicodeCharacter{2212}{−}
    \usepgfplotslibrary{groupplots,dateplot}
    \usetikzlibrary{patterns,shapes.arrows}
    \pgfplotsset{compat=newest}
    \def\BibTeX{{\rm B\kern-.05em{\sc i\kern-.025em b}\kern-.08em
        T\kern-.1667em\lower.7ex\hbox{E}\kern-.125emX}}
    \usepackage{float}

\usepackage[para,online,flushleft]{threeparttable}

\usepackage{graphicx,multirow}

    \makeatletter
    \newcommand{\linebreakand}
    {
        \end{@IEEEauthorhalign}
        \hfill\mbox{}\par
        \mbox{}\hfill\begin{@IEEEauthorhalign}
    }
    \makeatother
    
    \usepackage{arydshln}
    
\begin{document}
    
    \title{\MakeUppercase{Unsupervised Adaptive Deep Learning Method For BCI Motor Imagery Decoding}}

    \author
    {
        \IEEEauthorblockN{Yassine El Ouahidi, Giulia Lioi, Nicolas Farrugia, Bastien Pasdeloup and Vincent Gripon}
        \IEEEauthorblockA{\textit{IMT Atlantique} \\
                          \textit{Lab-STICC, UMR CNRS 6285} \\
                          F-29238 Brest, France \\
                          \emph{name}.\emph{surname}@imt-atlantique.fr}
    }

    \maketitle
    
    
    \begin{abstract}

    In the context of Brain-Computer Interfaces, we propose an adaptive method that reaches offline performance level while being usable online without requiring supervision.
    Interestingly, our method does not require retraining the model, as it consists in using a frozen efficient deep learning backbone while continuously realigning data, both at input and latent spaces, based on streaming observations.
    We demonstrate its efficiency for Motor Imagery brain decoding from electroencephalography data, considering challenging cross-subject scenarios.
    For reproducibility, we share the code of our experiments.
    \end{abstract}

    

    
\section{Introduction}
\label{sec:intro}

Brain-Computer Interfaces (BCIs) have gained considerable attention in neuroscience research due to their potential for various applications, from assisting people with disabilities to enhancing human-computer interaction~\cite{wolpaw2002brain}. Among BCI paradigms, Motor Imagery (MI) has attracted interest for its medical and societal benefits. MI-based BCIs translate mental imagery of motor movements into actionable commands or in a feedback for neuromotor rehabilitation~\cite{lioi2020multi}. As such, it opens up new ways of helping people with disabilities in their daily life.

Interpretation and classification of electroencephalography (EEG) signals are central to many BCI systems. Unfortunately, EEG exhibits low signal-to-noise ratio, non-stationarity and high variability between sessions and subjects~\cite{lotte2018review}. This make learning and classification challenging, especially when applying models trained offline to real-time BCI applications.

A common way to report performance in the literature consists in considering \emph{offline} settings where one can benefit from all experimental data for calibration and optimization. On the contrary, \emph{online} systems acquire data sequentially in real time, making it challenging to reach offline level performance. To overcome these difficulties, adaptive classifiers emerged as a promising solution. These classifiers update their parameters based on incoming EEG data, adapting to the evolving signal characteristics. One can distinguish two main types of adaptive classifiers: \textbf{supervised} or \textbf{unsupervised}~\cite{lotte2018review}. Supervised adaptive classifiers rely on labeled sequential data, while unsupervised classifiers do not require explicit labels, making them more suitable for realistic real-time BCI applications.

Recent advances in Deep Learning have improved the effectiveness of BCI systems. Indeed, Deep Learning models are particularly more efficient than classical Machine Learning alternatives for transferring knowledge between subjects. However, there are very few methods for adaptive Deep Learning classifiers, especially in the unsupervised setting.

In this work, we propose an unsupervised deep learning adaptive method for MI decoding and demonstrate its effectiveness using off-the-shelf benchmarks. Our method aims to bridge the gap between offline and online performance by dynamically updating normalization statistics throughout the considered Deep Learning architecture with each incoming data segment.
The main contributions of this paper are:
\begin{itemize}
    \item A deep learning method for unsupervised adaptive classification of MI signals, suitable for online setups;
    \item Experiments showing that our proposed method is competitive and achieves offline level performance, considering open source MI datasets and cross-subject scenarios;
    \item The code\footnote{\url{https://github.com/elouayas/eeg_adaptive}} to reproduce the experiments.
\end{itemize}

    
\section{Related Work}
\label{sec:related}


Machine Learning has significantly improved BCI systems, offering models that can effectively decode particular tasks. The MI paradigm is no exception, and remarkable progress has been made with the introduction of various EEG-specific BCI algorithms. In particular, methods using Riemannian geometry and Common Spatial Pattern (CSP) with automated feature selection have proven to be the most effective ones~\cite{lotte2018review}.

More recently, deep learning has introduced a promising way to exploit and transfer knowledge across sessions and subjects, and has emerged as a viable candidate solution to overcome the limitations of conventional Machine Learning methods. Multiple architectures have been proposed, including convolutional architectures with models such as Deep ConvNet, Shallow ConvNet~\cite{schirrmeister2017deep}, or EEGNet ~\cite{lawhern2018eegnet}. More recently EEG-SimpleConv~\cite{ouahidi2023strong}, a simple 1D-CNN, was proposed to serve as a robust baseline for MI decoding. In our study we use the EEG-SimpleConv architecture and training procedure, as it shows state-of-the-art performance.

The development of classifiers adapting to the dynamic nature of brain signals has been driven by the challenge of non-stationarity. In short, adaptive classifiers aim to maintain or improve classification performance over time despite variations in signal characteristics. Both supervised and unsupervised approaches to classifier adaptation have been proposed, each addressing various scenarios of BCI use. Supervised methods, which rely on labeled data to update the classifier, have shown promises but are limited by the availability of accurately labeled instances in real-time applications. On the contrary, unsupervised adaptive classifiers, which do not require labeled data, focus on updating the considered classifier to the evolving characteristics of the incoming data stream.

In the context of MI, unsupervised adaptation algorithms have been particularly highlighted for their potential in real-world BCI applications where the true intent of the user is not always explicitly known. Several methods applied to Linear Discriminant Analysis (LDA) and Gaussian mixture~\cite{blumberg2007adaptive,hasan2012hangman,liu2010improved} propose to estimate the class labels of new incoming samples before adapting the classifier. A simpler alternative~\cite{vidaurre2010toward} updating the bias only has also been applied to LDA.  
In Riemannian approaches, an unsupervised adaptive Minimum Distance to Mean (MDM) classifier~\cite{kumar2019towards} has been proposed to retrain the classifier after each prediction. Another adaptive Riemannian classifier~\cite{bougrain2021guidelines} does not require retraining but only recentering. As in~\cite{vidaurre2010toward,bougrain2021guidelines}, our method needs no retraining of the model. 

Not many adaptive approaches were proposed for Deep Learning classifiers. To the best of our knowledge, only a recent work~\cite{wimpff2023calibration} addresses this gap using various normalization techniques. We use this work as a comparison point.

    
\section{Methodology}
\label{sec:methodology}

Machine learning models, especially Deep Learning ones, perform best when the training and test sets belong to the same distribution. However, this premise faces significant challenges while working with EEG data, particularly in BCI applications. Indeed, due to the inherent variability of EEG signals across subjects and sessions, a mismatch in data distributions is common, leading to suboptimal performance of BCI models. To address this issue, our study introduces a novel approach aimed at minimizing distributional discrepancies, thereby enhancing the utility of pre-trained offline neural network for new, unseen data and facilitating their application in an online context.

Our proposed methodology consists in using an unsupervised adaptive Deep Learning classifier, characterized by its ability to adapt to new data without the need for retraining. This is achieved by dynamically updating the test data statistics when a new EEG trial arrives. The process involves two key steps: first, the incoming trial is aligned using updated Euclidean Alignment (EA)~\cite{he2019transfer} statistics; then, batch normalization layers statistics within the neural network's latent space are updated. This dual strategy of alignment and normalization is inspired by previous research demonstrating its effectiveness in MI decoding tasks~\cite{ouahidi2023strong}. By incorporating these adaptive elements into our model, we aim to bridge the distributional gaps between training and test data sets, thereby facilitating the use of Deep Learning models in real-time BCI applications.


\subsection{Euclidean Alignement (EA)}

EA enhances BCI models generalization by standardizing EEG data across different subjects and sessions into a domain-invariant space. For a given subject, it aligns each session's EEG trial data using the arithmetic mean of the covariance matrices. Through EA, data are transformed such that the mean covariance matrix of the trials is the identity matrix, significantly reducing dissimilarities across subjects and sessions, resulting in better homogeneity~\cite{he2019transfer}. EA is a data transformation step that has proven its worth in BCI and especially in MI decoding, and could be considered as commonly accepted in the offline literature~\cite{junqueira2024systematic}.
In our adaptive setup, EA plays a crucial role to  recalibrate the data we feed into our classifier in real time, without the need for labeled data.

Formally, let us consider a subject with $n$ trials, where each trial ${\bf X}_i \in \mathbb R^{C \times T}$ is composed of $C \times T$ samples with $C$ the number of channels and $T$ the number of time samples. Let us define:
\begin{equation}
{\bf \bar{R}}=\frac{1}{n} \sum_{i=1}^{n} {\bf X}_i {\bf X}^{T}_i 
\;.
\end{equation}
Alignment is performed by using the matrix square root of the arithmetic mean:
\begin{equation}
{\bf \tilde{X}}_i= {\bf \bar{R}}^{-\frac{1}{2}} {\bf X}_i
\:.
\end{equation}

EA is generally applied offline, as we often wait for all the trials from a given subject or session to become available to perform the alignment with the estimated matrix ${\bf \bar{R}}$. 
To adapt it to the online setting, we propose to recompute an updated version of the matrix $\bf \bar{R}$ each time a new trial arrives, performing the alignment ${\bf \tilde{X}}_i$ on the way.

\subsection{Batch normalization (BN) ``trick''}

BN is a widely used and effective technique in Deep Learning. BN consists in normalizing the activation within the neural network per batch. However, its behaviour during training and testing is different. During training, the running mean and variance are calculated and updated per batch. During evaluation, the learned and stored statistics from training are used to normalize the input. To adapt BN to online testing, we compute statistics directly from the test sets, rather than relying on stored training statistics. More specifically, we adopt an adaptive approach, and for each new incoming trial, we recompute the statistics with that trial and all previous ones from the same session.  
This allows session-specific statistics to be estimated and improves the ability of the model to adapt to individual users and non-stationarity.

\subsection{Warm-up buffer of data}
\label{sec:buffermethodo}

Our method relies on updating incoming test trials via normalization. However, to get good estimates of normalization statistics, it is necessary to have several trials. As a consequence, normalization of the first few trials may be inaccurate due to imprecise estimation, resulting in a poor initial performance of our adaptive classifier. To overcome this, we propose to  initially approximate those statistics using a buffer of data randomly sampled from the calibration set or from other subjects, depending on the evaluation settings (see Section~\ref{sec:evalsettings}). This buffer, together with incoming trials, are used to compute the statistics up to a certain point. After acquiring a large enough number of trials, we switch to only using session-specific trials to get reliable statistics. 

We choose to evaluate our method in different settings described in the next section, Cross-Subject and Cross-Subject with Fine-Tuning, and compare it to offline and online alternatives.

\subsection{Evaluation Settings}
\label{sec:evalsettings}

In this study, we evaluate the capabilities of our adaptive approach to transfer knowledge across domains. One of the main advantages of Deep Learning approaches over Machine Learning approaches lies in their ability to enable transfer learning~\cite{ouahidi2023strong}. 
More specifically, it allows the use of data from multiple subjects simultaneously to transfer the learned knowledge to a new subject. We consider the following scenarios:
\begin{itemize}
    \item \textbf{Cross-Subject.} In this scenario, the model is trained on data from all but one subject, and its performance is evaluated on data from the omitted subject;
    \item \textbf{Cross-Subject with Fine-Tuning.} Here we pre-train the model on data from all but one of the subjects, then we fine-tune on a part of the data from the omitted subject. We then evaluate the performance of the model on the rest of the data from the omitted subject.
\end{itemize}

\subsection{Offline, Online and Adaptive Setups}

In the offline evaluation setup for MI classification, we consider a scenario in which we want to classify an ensemble of recorded trials, with access to all of them at any time, in an unsupervised way (without having access to their labels). In contrast, in the online evaluation setup, we make predictions on individual trials, one at a time. Consequently, we cannot use the rest of data from the test set for normalization or alignment. The online setup simulates a strict real-life application of a classifier.
Due to the varying nature of EEG signals, using more trials from the same session greatly improves classification performance. In \cite{ouahidi2023strong}, we observe a gap of about $15\%$ on a 4-class classification task on a MI benchmark dataset.
The adaptive setup, which lies between the online and offline setups, simulates a realistic but less strict use of a classifier, where we have access to the incoming trial and the previous ones. In this setup, we make predictions on individual trials using also previously classified trials to update normalization statistics. Importantly, the proposed normalization techniques do not need trial labels, which makes the method unsupervised.

\section{Experiments \& results}
\label{sec:experiments}

\subsection{Datasets and preprocessing}
To validate our method, we consider two open access datasets, 
a small-scale dataset, BNCI~\cite{tangermann2012review} (the IIa dataset from the BCI competition) and the Large dataset from~\cite{dreyer2023large}, which contains 9 times more subjects than BNCI. In our experiments we also consider a subset of the BNCI dataset, BNCI2, which contains EEG data belonging to only two classes (Right and Left MI) in order to allow a fair comparison with the Large dataset.

Each dataset has a unique recording setup with different devices and configurations, as detailed in Table~\ref{tab:datasets}. In our experiments, we consider signals starting from the cue indicating the task to be imagined. 

\begin{table}[H]
\center
\caption{MI datasets considered. L = Left hand, R = Right hand, F = Feet, T = Tongue.}
\begin{tabular}{l|llll}
\hline
Dataset                      & BNCI & BNCI2 & Large \\ \hline
Considered subjects          &   9     &  9   & 85   \\
Sessions per subject         &   2     &  2   &  5   \\ 
Trials per session           &   288   & 144  &  40  \\ 
Total trials                 &   5184  & 2592 & 17000 \\ 
Classes                      & L/R/F/T &  L/R & L/R    \\ 
EEG electrodes               &   22     & 22    &   27    \\ 
Sampling frequency (Hz)      &  250 & 250  &  512   \\ 
Trial duration (s)           &   4     &   4   &   4.5       \\ \hline
\end{tabular}
 \label{tab:datasets}
\end{table}

We used the very minimal preprocessing steps suggested in ~\cite{ouahidi2023strong,el2023spatial}. The idea is to take advantage of the ability of Deep Learning models to work with raw data and leave them as much latitude as possible to learn: we apply a high-pass filter at 0.5 Hz, as recommended in~\cite{Delorme2023}, and resample to a lower sampling frequency to speed up the model's 
inference, making sure to keep frequencies up to 40-50 Hz.

We follow the same training procedure as described in~\cite{ouahidi2023strong}. 
For the cross-subject paradigm, one backbone per subject is trained for each run. To train this backbone, we use the data from all subjects except the evaluated subject.
In the Cross-Subject with Fine-Tuning scenario, we reuse the backbone previously trained in the Cross-Subject paradigm and calibrate it on the evaluated subject. To perform the calibration, we use the sessions as follow: for BNCI, the first session in the omitted subject is used for fine-tuning and the second for evaluation; for the Large database, we use the first two sessions for fine-tuning and the last three for evaluation.

\subsection{Main Results}

\begin{figure*}[h!]
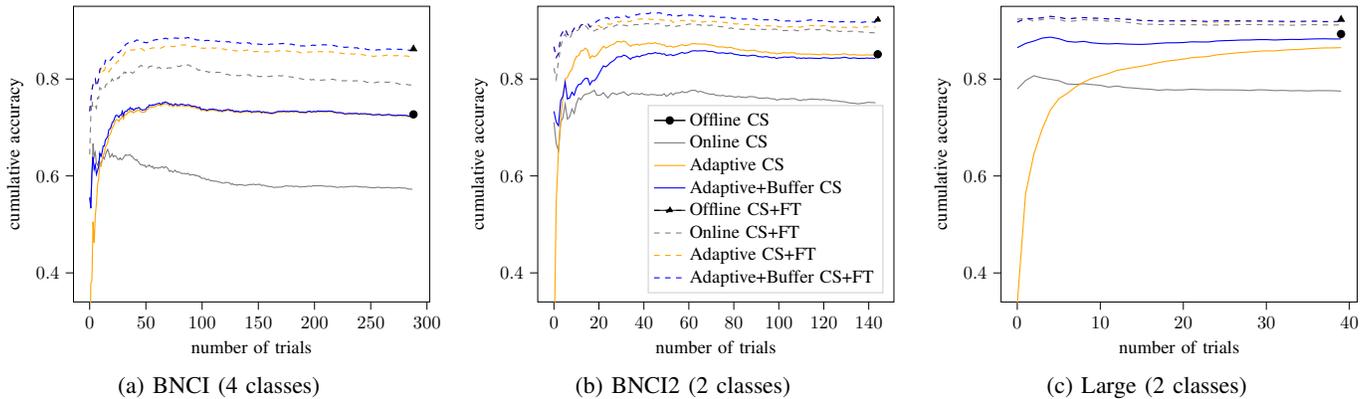

    \centering
    \begin{subfigure}{0.32\linewidth}
        \centering
        \resizebox{!}{4.7cm}{\input{figures/bnci.tex}}
        \caption{BNCI (4 classes)}
    \end{subfigure}
    \hfill
    \begin{subfigure}{0.32\linewidth}
        \centering
        \resizebox{!}{4.7cm}{\input{figures/bnci2.tex}}
        \caption{BNCI2 (2 classes)}
    \end{subfigure}
    \hfill
    \begin{subfigure}{0.32\linewidth}
        \centering
        \resizebox{!}{4.7cm}{
\begin{tikzpicture}

\definecolor{darkgray176}{RGB}{176,176,176}
\definecolor{gray}{RGB}{128,128,128}
\definecolor{lightgray204}{RGB}{204,204,204}
\definecolor{orange}{RGB}{255,165,0}

\begin{axis}[
legend cell align={left},
legend style={
  fill opacity=0.8,
  draw opacity=1,
  text opacity=1,
  at={(0.97,0.03)},
  anchor=south east,
  draw=lightgray204
},
tick align=outside,
tick pos=left,
x grid style={darkgray176},
xlabel={number of trials},
xmin=-1.95, xmax=40.95,
xtick style={color=black},
y grid style={darkgray176},
ylabel={cumulative accuracy},
ymin=0.34, ymax=0.95,
ytick style={color=black}
]
\addplot [draw=black, draw=none, fill=black, mark=*]
table{%
x  y
39 0.893
};
\addplot [draw=black, draw=none, fill=black, mark=triangle*]
table{%
x  y
39 0.923
};
\addplot [semithick, gray]
table {%
0 0.779411764705882
1 0.797058823529412
2 0.806862745098039
3 0.802205882352941
4 0.799411764705882
5 0.79656862745098
6 0.789915966386555
7 0.788970588235294
8 0.790522875816993
9 0.788529411764706
10 0.787165775401069
11 0.783333333333333
12 0.785746606334842
13 0.782563025210084
14 0.78156862745098
15 0.779963235294117
16 0.779065743944637
17 0.776960784313725
18 0.778173374613003
19 0.777352941176471
20 0.777731092436975
21 0.779010695187166
22 0.778260869565217
23 0.777696078431373
24 0.777882352941176
25 0.777601809954751
26 0.77755991285403
27 0.777310924369748
28 0.776369168356998
29 0.777843137254902
30 0.777514231499051
31 0.776194852941176
32 0.776648841354724
33 0.775778546712803
34 0.776218487394958
35 0.775490196078431
36 0.776152623211447
37 0.776160990712074
38 0.775942684766214
39 0.775147058823529
};
\addplot [semithick, orange]
table {%
0 0.335294117647059
1 0.563235294117647
2 0.645098039215686
3 0.697058823529412
4 0.736470588235294
5 0.759313725490196
6 0.770168067226891
7 0.783455882352941
8 0.793790849673203
9 0.801470588235294
10 0.806417112299465
11 0.811519607843137
12 0.817873303167421
13 0.820798319327731
14 0.823921568627451
15 0.826654411764706
16 0.830103806228374
17 0.833823529411765
18 0.836996904024768
19 0.838970588235294
20 0.841176470588235
21 0.844251336898396
22 0.84539641943734
23 0.84656862745098
24 0.849882352941177
25 0.851696832579186
26 0.853159041394336
27 0.854936974789916
28 0.855882352941177
29 0.857450980392157
30 0.858254269449716
31 0.858180147058824
32 0.859536541889483
33 0.860726643598616
34 0.86109243697479
35 0.8625
36 0.863513513513513
37 0.864241486068111
38 0.864328808446456
39 0.864852941176471
};
\addplot [semithick, blue]
table {%
0 0.864705882352941
1 0.873529411764706
2 0.87843137254902
3 0.884558823529412
4 0.886470588235294
5 0.883333333333333
6 0.877731092436975
7 0.876102941176471
8 0.877777777777778
9 0.874705882352941
10 0.873262032085562
11 0.872549019607843
12 0.873303167420815
13 0.872268907563025
14 0.871764705882353
15 0.871507352941177
16 0.872318339100346
17 0.873692810457516
18 0.874767801857585
19 0.87485294117647
20 0.875350140056022
21 0.876871657754011
22 0.876598465473146
23 0.876470588235294
24 0.878588235294118
25 0.879298642533937
26 0.879738562091503
27 0.880567226890756
28 0.880628803245436
29 0.881372549019608
30 0.881404174573055
31 0.880606617647059
32 0.881283422459893
33 0.881833910034602
34 0.881596638655462
35 0.882434640522876
36 0.882909379968204
37 0.88312693498452
38 0.882730015082956
39 0.882794117647059
};
\addplot [semithick, gray, dashed]
table {%
0 0.923529411764706
1 0.925
2 0.919607843137255
3 0.922794117647059
4 0.925294117647059
5 0.923039215686274
6 0.918487394957983
7 0.920220588235294
8 0.922549019607843
9 0.921470588235294
10 0.920588235294118
11 0.919362745098039
12 0.920135746606335
13 0.917436974789916
14 0.915686274509804
15 0.913051470588235
16 0.913148788927336
17 0.912418300653595
18 0.912383900928792
19 0.912941176470588
20 0.911904761904762
21 0.912032085561497
22 0.912148337595908
23 0.911274509803921
24 0.911411764705882
25 0.911990950226244
26 0.911764705882353
27 0.91281512605042
28 0.912373225152129
29 0.91235294117647
30 0.912049335863377
31 0.911856617647058
32 0.911675579322638
33 0.911072664359861
34 0.911428571428571
35 0.912336601307189
36 0.912162162162162
37 0.911687306501548
38 0.911463046757164
39 0.911838235294117
};
\addplot [semithick, orange, dashed]
table {%
0 0.917647058823529
1 0.925
2 0.924509803921569
3 0.926470588235294
4 0.929411764705882
5 0.927450980392157
6 0.923949579831933
7 0.925735294117647
8 0.926797385620915
9 0.925588235294118
10 0.925401069518717
11 0.924019607843137
12 0.924886877828054
13 0.923529411764706
14 0.922745098039216
15 0.921139705882353
16 0.921107266435986
17 0.919934640522876
18 0.919659442724458
19 0.919558823529412
20 0.918487394957983
21 0.918983957219251
22 0.918158567774936
23 0.91703431372549
24 0.917529411764706
25 0.918212669683258
26 0.918082788671024
27 0.91890756302521
28 0.918661257606491
29 0.918921568627451
30 0.918406072106262
31 0.918106617647059
32 0.917825311942959
33 0.917474048442906
34 0.917899159663865
35 0.918464052287582
36 0.918600953895071
37 0.917879256965944
38 0.917797888386124
39 0.918529411764706
};
\addplot [semithick, blue, dashed]
table {%
0 0.917647058823529
1 0.925
2 0.924509803921569
3 0.926470588235294
4 0.929411764705882
5 0.927450980392157
6 0.923949579831933
7 0.925735294117647
8 0.926797385620915
9 0.925588235294118
10 0.925401069518717
11 0.924019607843137
12 0.924886877828054
13 0.923529411764706
14 0.922745098039216
15 0.920588235294118
16 0.920415224913495
17 0.919444444444445
18 0.920278637770898
19 0.920147058823529
20 0.9203081232493
21 0.920855614973262
22 0.919948849104859
23 0.91875
24 0.920235294117647
25 0.92024886877828
26 0.920043572984749
27 0.920378151260504
28 0.919878296146045
29 0.920098039215686
30 0.919639468690702
31 0.919485294117647
32 0.919162210338681
33 0.918944636678201
34 0.919075630252101
35 0.91952614379085
36 0.919634340222575
37 0.918653250773994
38 0.918702865761689
39 0.919264705882353
};
\end{axis}

\end{tikzpicture}}
        \caption{Large (2 classes)}
    \end{subfigure}
    \caption{Adaptive classifier performance compared to online and offline baselines in Cross-Subject (CS) and Cross-Subject Fine-Tuning (CS+FT) evaluations.}
    \label{fig:mainresults}
\end{figure*}

Results are shown in Table~\ref{fig:perf}. We observe that our adaptive method is as effective as the offline baseline and highly outperforms the online baseline. This has been observed for every dataset and for every evaluation paradigm considered.
A notable point is that we can observe that the results of the two-class datasets, BNCI2 and Large, are similar, which is unsurprising considering the similarity of tasks and setups.

The line termed ``Soft K-means'' in Table~\ref{fig:perf} refers to an alternative method of performing the classification in our adaptive method. The normalisation steps are the same as described previously. However, we perform a Soft K-means clustering on the predicted class probabilities at the output of the model. The idea is to exploit the incoming trials during the evaluation session to refine and obtain better decision frontiers for the classification compared with the fixed classification layer. However, we can see that the result is similar and does not add significant value. This could be explained by the fact that the decision boundary is already adjusted by the normalisations, in particular the internal normalisation using the BN-trick.

Our method is able to achieve near-offline performance with only 10 to 20 trials (depending on the number of classes and the scoring paradigm). This corresponds to a EEG acquisition of about 30 to 60 seconds in a MI setup. Figure~\ref{fig:mainresults} provides more details on the behaviour of our method over time. In this Figure, we depict the cumulative accuracy over the incoming trials, \emph{i.e.}, the average accuracy up to this trial. We can observe the rapid convergence of the method to the offline baseline performance, together with its superiority over the online baseline at all given time steps.

\begin{table}[h]
\caption{Comparison of accuracies (\%) of various methods.}
\label{fig:perf}
\center
\begin{tabular}{llll}
\hline
                  & BNCI & BNCI2 & Large   \\ \hline
                & \multicolumn{3}{c}{Cross-Subject}   \\ \hline
 Online     & 56.9 & 75.1  & 77.5    \\ 
 Adaptive        & 72.1 & 84.9  & 86.5    \\ 
 Adaptive + Buffer & 72.2 & 84.2  & 88.3    \\ 
 Offline     & 72.7 & 85.1  & 89.3    \\ \hline
                & \multicolumn{3}{c}{Cross-Subject + Fine-Tuning}   \\ \hline
Online      & 78.8 & 89.5  & 91.2    \\ 
Adaptive           & 84.6 & 90.7  & 91.9    \\ 
 Adaptive + Buffer   & 85.9 & 91.7  & 91.9    \\ 
Adaptive + Soft K-means     & 84.6 & 90.6  & 91.8    \\ 
Offline       & 86.2 & 92.2  & 92.3     \\ \hline

\end{tabular}
\end{table}

\begin{table}[h]
\caption{Comparison to other unsupervised adaptive methods on BNCI Cross-Subject evaluation.}
\center
\begin{threeparttable}
\begin{tabular}{llll}
\hline
                  & RCT~\cite{bougrain2021guidelines}\tnote{1} & Wimpff et al.~\cite{wimpff2023calibration} & Ours   \\ \hline
Accuracy (\%)         & 44.6 & 67.3 & \textbf{72.2}    \\  \hline

\end{tabular}
\begin{tablenotes}
\item[1] Results reproduced by us

  \end{tablenotes}
\end{threeparttable}
\label{tab:sota}
\end{table}

In Table~\ref{tab:sota} we can see that in the Cross-Subject setting, our method outperforms the other Deep Learning method~\cite{wimpff2023calibration}, by 6\%. Since they have many similarities, this shows the importance of using a good backbone with the right training routine. 
Our method outperforms RCT even more by almost 30\%. However, it should be pointed out that RCT was not designed to allow transfer between subjects, and was not meant to be used in a Cross-Subject scenario. We can assume that it would be way more efficient in a Within-Subject scenario.

\subsection{Additional experiments}

\subsubsection{Warm-up buffer of trials}


We investigated the number of early trials for which using a warm-up buffer is beneficial. We observed that this had a marginal impact on the overall score, and found an optimum at around 10 trials. We also investigated the optimal buffer size. We found that 40 trials is enough, and we did not notice benefits while further increasing the size.

The conclusions from these experiments are that in all the setups and datasets, the addition of a buffer is harmless to performance and generally provides a large gain at the beginning of the evaluation, when the first trials are classified. This is shown in Figure~\ref{fig:mainresults} where Adaptive+Buffer (blue) outperforms Adaptive (orange) for the left part with few trials (with the exception of Cross Subject, BNCI2), and at the same performance level or above in the long run. We therefore recommend using a buffer to maximize performance and ensure reliable classification from the start.


Other sorts of buffers have been tried. For example, a fixed size sliding-window buffer in order to use only the last trials to normalize, with the rationale that there might be variations in the signals within long sessions. However, this did not show significantly better results, at the cost of more complexity.

\subsubsection{Hypothesis on user fatigue}

\begin{figure}[h!]
    \centering
    \begin{subfigure}{0.8\linewidth}
        \centering
        \resizebox{\linewidth}{!}{\input{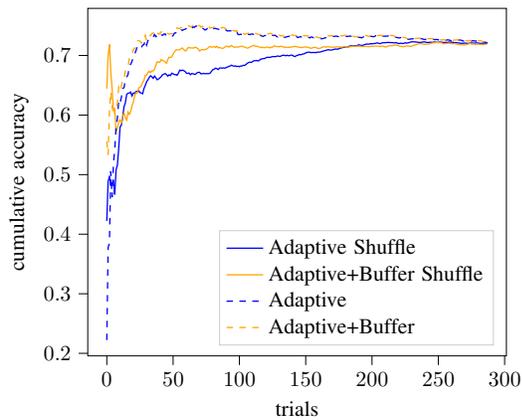}}
        \caption{Cross-Subject}
    \end{subfigure}
        \\~\\
    \begin{subfigure}{0.8\linewidth}
        \centering
        \resizebox{\linewidth}{!}{\input{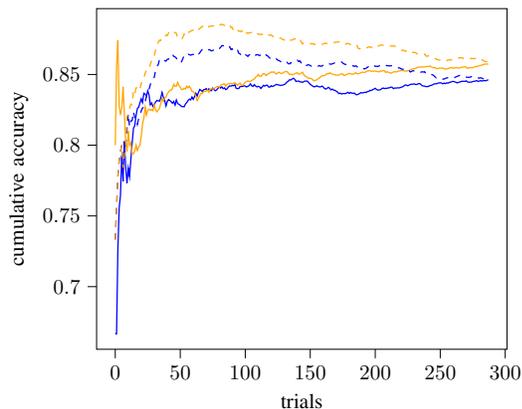}}
        \caption{Cross-Subject + Fine-Tuning}
    \end{subfigure}
    \caption{Impact of shuffling the data to test the hypothesis on user fatigue, on the BNCI dataset.}
    \label{fig:shuffle}
\end{figure}

We can observe that classification performance of our adaptive method seems to decline over time. This can be seen in the dotted curves of Figure~\ref{fig:shuffle}, which decreases after a certain point. We hypothesize that subjects experience fatigue at the end of the session, with poorer concentration during the final trials. BCI sessions have been previously shown to be mentally exhausting~\cite{shenoy2006towards}, which is further enhanced by the long duration of BNCI sessions (288 trials).

To evaluate this hypothesis, we shuffled each session trials so that our adaptive method was applied to the trials in a random order rather than a chronological order. By doing so, we can see that the curves are strictly increasing. Also, the shuffled and unshuffled trials achieve the same final performance, but with different trends. This experience supports our hypothesis on user fatigue affecting performance.

\section{Conclusions}
\label{sec:conclusions}

We have proposed a method for the adaptive use of BCI deep learning backbones. Our method can be used in real time and does not require the model to be re-trained. It consists in updating normalizations at both input and latent spaces. We have demonstrated its effectiveness on several MI datasets.
The method can be particularly useful for faster calibration of a system in a few-shot setup. The effectiveness of the method remains to be demonstrated in a real online acquisition study.

    \bibliographystyle{IEEEtran}
    \bibliography{4_references}
\end{document}